\documentclass{mn2e}
\usepackage{epsfig}

\title[A Census of the Carina Nebula]{A Census of the Carina
Nebula. I: Cumulative Energy Input from Massive Stars}

\author[N.\ Smith]{Nathan Smith\thanks{Hubble Fellow;
nathans@casa.colorado.edu} \\ Center for Astrophysics and Space
Astronomy, University of Colorado, 389 UCB, Boulder, CO 80309, USA}

\date{Accepted 0000, Received 0000, in original form 0000}
\pagerange{\pageref{firstpage}--\pageref{lastpage}} \pubyear{2002}

\def\arcdeg{\degr}
\begin{document}
\label{firstpage}
\maketitle
\begin{abstract}

The Carina Nebula (NGC~3372) is our richest nearby laboratory in which
to study feedback through UV radiation and stellar winds from very
massive stars during the formation of an OB association, at an early
phase in the evolution of the surrounding proto-superbubble before
supernova explosions have influenced the environment.  This feedback
is triggering successive generations of new star formation around the
periphery of the nebula, while simultaneously evaporating the gas and
dust reservoirs out of which young stars are trying to accrete
material.  This paper takes inventory of the combined effect from all
the known massive stars that power the Carina Nebula through their
total ionizing flux and integrated mechanical energy from their
stellar winds. Carina is close enough and accessible enough that
spectral types for individual stars are available, and many close
binary and multiple systems have recently been spatially resolved, so
that one can simply add them.  Adopting values from the literature for
corresponding spectral types, the present-day total ionizing photon
luminosity produced by the 65 O stars and 3 WNL stars in Carina is
Q$_H\simeq$10$^{51}$ s$^{-1}$, the total bolometric luminosity of all
stars earlier than B2 is 2.5$\times$10$^7$ $L_{\odot}$, and the total
mechanical luminosity of stellar winds is $L_{SW}\simeq$10$^5$
$L_{\odot}$.  The total Q$_H$ was about 25\% higher when
$\eta$~Carinae was on the main sequence, before it and its companion
were surrounded by its obscuring dust shell; for the first 3 Myr, the
net ionizing flux of the 70 O stars in Carina was about 150 times
greater than in the Orion Nebula.  About 400-500 $M_{\odot}$ has been
contributed to the H~{\sc ii} region by stellar wind mass loss during
the past 3 Myr.  Values for Q$_H$ and $L_{SW}$ are also given for the
individual clusters Tr14, 15, and 16, and Bo10 and 11, which are more
relevant on smaller spatial scales than the total values for the whole
nebula.
 
\end{abstract}


\section{INTRODUCTION}

Feedback from young massive stars may play an integral role in star
and planet formation.  Most stars are born in the vicinity of the
hottest and most massive stars spawned only from giant molecular
clouds, and the effects of feedback from these massive stars cannot be
studied in nearby quiescent regions of star formation.  Even the
nearest H~{\sc ii} region, the Orion Nebula, may not be representative
of the extreme environments where most stars are born, since it is
dominated by just a single O6 dwarf.

A much more extreme collection of stars can be found in the Carina
Nebula (NGC~3372).  Its star clusters are almost as spectacular as
those of 30~Dor in the LMC (Massey \& Hunter 1998), or objects in our
own Galaxy like the Arches cluster near the Galactic center (Najarro
et al.\ 2004; Figer et al.\ 1999, 2002) and NGC~3603 (Moffat et al.\
2002).  However, these other regions are too distant for detailed
studies of small-scale phenomena like irradiated protoplanetary disks
and jets, and their study is hampered by considerably more extinction.
The low extinction toward Carina combined with its proximity and rich
nebular content provide a worthwhile trade-off for its less
concentrated star clusters.  While this may be a liability for
investigating the upper end of the initial mass function, the fact
that Carina's stars are more of a loose aggregate may be an advantage
for studying feedback and triggered star formation.  Carina provides a
snapshot of an OB association in the making, which may be more
representative of the environments in which most stars form than are
bound super star clusters.  It may also provide an early analog of the
local Galactic environment that evolved into Gould's Belt.

Recent studies suggest that at the present epoch there exists an upper
mass limit to the most massive stars of $\sim$150 M$_{\odot}$ (Figer
2005, Weidner \& Kroupa 2004, Kroupa \& Weidner 2005; Oey \& Clarke
2005).  While the star clusters that power Carina may not be as dense
as the Arches cluster or NGC~3603, Carina does contain several
examples of stars that probably began their lives with 100--150
M$_{\odot}$.  Among these are $\eta$~Carinae, the prototypical O2
supergiant HD~93129A, three late-type hydrogen-rich Wolf-Rayet stars
(WNL stars), and the remaining original members of the O3 spectral
class for which this spectral type was first introduced (Walborn 1973;
although see Walborn et al.\ 2002).  For these colossal stars,
lifetimes of $\sim$3 Myr before they explode as supernovae (SNe) are
shorter than the time it takes to clear away their natal molecular
material.  Consequently, we already have a situation in Carina where
the most massive members like $\eta$~Car are approaching their
imminent demise while new stars are being born from dense molecular
gas only 5--20 pc away.  In the next 1--2 Myr, there will be several
very energetic SNe in the Carina Nebula, which will carve out an even
larger cavity in the ISM and form a giant superbubble in the Galactic
plane, and may pollute protoplanetary disks with nuclear-processed
ejecta.  In the mean time, studying this rich region provides a
snapshot of a young proto-superbubble (Smith et al.\ 2000) energized
only by UV radiation and stellar winds, immediately before the
disruptive SN-dominated phase.

In just the past decade, the Carina Nebula has been recognized as a
hotbed of ongoing active star formation.  It provides a laboratory to
study several star formation phenomena in great detail, all of which
have recently been identified in this region: (1) evaporating
protoplanetary disks (the so-called ``proplyds''), small cometary
clouds, or globules (Smith et al.\ 2003$a$; Smith 2002$b$; Brooks et
al.\ 2000, 2006; Cox \& Bronfman 1995), (2) the erosion of large dust
pillars and the triggering of a second generation of embedded star
formation within them (Smith et al.\ 2000, 2005; Rathborne et al.\
2004; Megeath et al.\ 1996), (3) irradiated Herbig-Haro (HH) jets that
are a signpost of active accretion (Smith et al.\ 2004), (4)
photodissociation regions (PDRs) on the surfaces of molecular clouds
across the region (Brooks et al.\ 2003, 2006, 1998; Rathborne et al.\
2002; Smith et al.\ 2000; Mizutani et al.\ 2004), and on the largest
scales, (5) the early formation of a fledgeling superbubble (Smith et
al.\ 2000).  These phenomena trace a second generation of stars,
perhaps triggered by feedback from the first generation clusters.
Carina is near enough that an upcoming program with the {\it Hubble
Space Telescope} will undertake the same types of detailed studies of
feedback that have been done in Orion at visual and near-IR
wavelengths, but sampling a more extreme environment.

All of these phenomena respond to the cumulative UV radiation and/or
stellar winds from star clusters that power the H~{\sc ii} region, and
a census of this energy input is essential in order to understand the
relationship between them.  Estimating the total stellar energy input
in Carina is a more difficult problem than in smaller H~{\sc ii}
regions like Orion, where a single star dominates the photoevaporation
of proplyds and the irradiation of HH jets.  Instead, Carina has
dozens of massive O stars over many parsecs.  Although the remarkable
stellar content of the Carina Nebula has been discussed many times in
the literature (e.g., Walborn 1995, 2002, 2005; Feinstein 1995; Massey
\& Johnson 1993), there is no complete and up-to-date census of the
massive stars, and especially there has been no estimate of their {\it
collective} UV radiation and mechanical (wind) luminosity that powers
the region.  This is the main purpose of the present investigation.  A
second paper in this series will compare the total energy input from
stars with the apparent energy budget inferred from observations of
the surrounding Carina Nebula.

\section{ADOPTED PARAMETERS}

The basic strategy in this paper will be to use the observed spectral
type of each massive star (spectral types as late as B2 will be
considered, although their influence is minimal) to assign a
corresponding value for the hydrogen-ionizing photon luminosity $Q_H$,
as well as the mass-loss rate $\dot{M}$ and terminal wind speed
V$_{\infty}$, and hence, the wind mechanical luminosity
$L_{SW}$=(1/2)$\dot{M}$V$_{\infty}^2$.  Table 1 lists the values that
will be adopted for stars later than O3.  In Table 1, values of log L
and $Q_H$ for each spectral type are taken from the ``observed''
temperature-scale models of Martins, Schaerer, \& Hillier (2005).
Values for $\dot{M}$ are taken from Repolust, Puls, \& Herrero (2004),
and terminal velocities for each spectral type are taken by a rough
fit to the average values listed by Prinja, Barlow, \& Howarth
(1990).\footnote{Given the spread in terminal velocities listed by
Prinja et al.\ 1990, we adopted the same velocities for all three
luminosity classes in Table 1.  This is a reasonable approximation for
the earliest spectral types that dominate the energy budget.}

The stellar content of the Carina Nebula is truly remarkable and
accessible, and several of the more spectacular stars have received
enough attention that the desired parameters have been derived
individually for them.  In some cases, when noted below, individual
studies may supersede the average values adopted in Table 1.  Many of
these stars are, in fact, benchmarks for their spectral types in the
three papers just mentioned.  In addition, the three WNL stars in
Carina are of particular interest, as are HD~93129A and $\eta$~Car
(when it was on the main sequence), since these are the most luminous
sources that have the most influence.  The specific stellar content of
the nebula is described in more detail in the following section.

%
%
\begin{table*}\begin{minipage}{3.4in}
\caption{Adopted Parameters for Each Spectral Type}\scriptsize
\begin{tabular}{@{}lcccccc}\hline\hline
Spect. &Lum. &log L &log $Q_H$ &$\dot{M}$ &V$_{\infty}$ &$L_{SW}$ \\
Type &Class &(L$_{\odot}$) &(s$^{-1}$) &(10$^{-6}$ M$_{\odot}$ yr$^{-1}$) 
            &(km s$^{-1}$) &(L$_{\odot}$) \\ \hline
O3	&V	&5.84	&49.64	&3.1	&3160	&2450	\\
O3.5	&V	&5.76	&49.54	&2.5	&3080	&1870	\\
O4	&V	&5.67	&49.44	&2.0	&2990	&1410	\\
O4.5	&V	&5.58	&49.33	&1.6	&2870	&1040	\\
O5	&V	&5.49	&49.22	&1.3	&2760	&780	\\
O5.5	&V	&5.41	&49.10	&1.0	&2650	&550	\\
O6	&V	&5.32	&48.99	&0.8	&2560	&410	\\
O6.5	&V	&5.23	&48.88	&0.6	&2460	&290	\\
O7	&V	&5.14	&48.75	&0.4	&2320	&170	\\
O7.5	&V	&5.05	&48.61	&0.35	&2210	&135	\\
O8	&V	&4.96	&48.44	&0.26	&2100	&91	\\
O8.5	&V	&4.86	&48.27	&0.22	&1960	&67	\\
O9	&V	&4.77	&48.06	&0.18	&1820	&47	\\
O9.5	&V	&4.68	&47.88	&0.15	&1660	&33	\\
B0	&V	&4.57	&47.70	&0.12	&1510	&22	\\
B0.5	&V	&4.47	&47.50	&0.10	&1460	&17	\\
B1	&V	&4.37	&47.28	&0.08	&1180	&8.8	\\
B1.5	&V	&4.28	&47.05	&0.06	&960	&4.4	\\
B2	&V	&4.19	&46.80	&0.05	&750	&2.2	\\ \hline
O3	&III	&5.96	&49.77	&6.5	&3160	&5130	\\
O3.5	&III	&5.91	&49.71	&5.5	&3080	&4120	\\
O4	&III	&5.85	&49.64	&4.7	&2990	&3320	\\
O4.5	&III	&5.79	&49.56	&4.0	&2870	&2600	\\
O5	&III	&5.73	&49.48	&3.4	&2760	&2050	\\
O5.5	&III	&5.67	&49.40	&2.7	&2650	&1500	\\
O6	&III	&5.61	&49.32	&2.2	&2560	&1140	\\
O6.5	&III	&5.54	&49.23	&1.8	&2460	&860	\\
O7	&III	&5.48	&49.13	&1.5	&2320	&640	\\
O7.5	&III	&5.42	&49.01	&1.2	&2210	&460	\\
O8	&III	&5.35	&48.88	&0.8	&2100	&280	\\
O8.5	&III	&5.28	&48.75	&0.7	&1960	&210	\\
O9	&III	&5.21	&48.65	&0.5	&1820	&130	\\
O9.5	&III	&5.15	&48.42	&0.4	&1660	&87	\\
B0	&III	&5.08	&48.28	&0.3	&1510	&54	\\
B0.5	&III	&5.00	&48.10	&0.25	&1460	&42	\\
B1	&III	&4.93	&47.90	&0.20	&1180	&22	\\
B1.5	&III	&4.86	&47.68	&0.15	&960	&11	\\
B2	&III	&4.78	&47.44	&0.11	&750	&4.9	\\ \hline
O3	&I	&5.99	&49.78	&13.0	&3160	&10300	\\
O3.5	&I	&5.96	&49.74	&11.3	&3080	&8470	\\
O4	&I	&5.93	&49.70	&10.0	&2990	&7060	\\
O4.5	&I	&5.90	&49.66	&8.9	&2870	&5790	\\
O5	&I	&5.87	&49.62	&8.0	&2760	&4810	\\
O5.5	&I	&5.84	&49.58	&7.1	&2650	&3940	\\
O6	&I	&5.81	&49.52	&6.4	&2560	&3310	\\
O6.5	&I	&5.78	&49.46	&5.7	&2460	&2730	\\
O7	&I	&5.75	&49.41	&5.0	&2320	&2130	\\
O7.5	&I	&7.72	&49.31	&4.3	&2210	&1660	\\
O8	&I	&5.68	&49.25	&3.6	&1960	&1090	\\
O8.5	&I	&5.65	&49.19	&3.0	&1820	&790	\\
O9	&I	&5.61	&49.11	&2.5	&1660	&540	\\
O9.5	&I	&5.57	&49.00	&1.9	&1510	&340	\\
%
\hline
\end{tabular}

References. --- L, $Q_H$: Martins et al.\ (2005); V$_{\infty}$: Prinja
et al.\ (1990); $\dot{M}$: Repolust et al.\ (2004). L and $Q_H$ for B
stars were extrapolated from Martins et al.\ (2005) following Crowther
(2005).
\end{minipage}
\end{table*}

%
%
\begin{table*}\begin{minipage}{4.25in}
\caption{Stellar Content of Trumpler 16 and Collinder 228}
\scriptsize
\begin{tabular}{@{}llccccc}\hline\hline
 Name &Sp. &log L &log $Q_H$ &$\dot{M}$ &V$_{\infty}$ &$L_{SW}$ \\
 \ &Type &(L$_{\odot}$) &(s$^{-1}$) &(10$^{-6}$ M$_{\odot}$ yr$^{-1}$) 
         &(km s$^{-1}$) &(L$_{\odot}$) \\ \hline
$\eta$~Car-now	&LBV		&6.67	&...	&1000	&600	&28700	\\
$\eta$~Car-MS	&O2~If~(?)	&6.57	&50.36	&8.32	&3200	&6780	\\
$\eta$~Car-B	&O5~V~(?)	&5.49	&49.2	&1.3	&2760	&780	\\
HD~93162	&WN6ha		&6.22	&50.0	&10.5	&2480	&5140	\\
HD~93131	&WN6ha		&5.94	&49.7	&13.8	&2160	&5130	\\
HD~92740A	&WN7ha		&6.01	&49.8	&14.5	&1785	&3680	\\
HD~92740B	&O8-9~V		&4.86	&48.27	&0.22	&1960	&67	\\
HD~93205A	&O3.5~V((f))	&5.76	&49.54	&2.5	&3080	&1870	\\
HD~93205B	&O8~V		&4.96	&48.44	&0.26	&2100	&91	\\
HD~93250	&O3.5~V((f))	&5.76	&49.54	&2.5	&3080	&1870	\\
HDE~303308	&O4~V((f))	&5.67	&49.44	&2.0	&2990	&1410	\\
HD~93204	&O5~V((f))	&5.49	&49.22	&1.3	&2760	&780	\\
Tr16-244	&O4~If		&5.93	&49.70	&10.0	&2990	&7060	\\
CPD~-59~2600	&O6~V((f))	&5.32	&48.99	&0.8	&2560	&410	\\
CPD~-59~2603A	&O7~V((f))	&5.14	&48.75	&0.4	&2320	&170	\\
CPD~-59~2603B	&O9.5~V		&4.68	&47.88	&0.15	&1660	&33	\\
CPD~-59~2603C	&B0.2~V		&4.57	&47.70	&0.12	&1510	&22	\\
CPD~-59~2628A	&O9.5~V		&4.68	&47.88	&0.15	&1660	&33	\\
CPD~-59~2628B	&B0.3~V		&4.57	&47.70	&0.12	&1510	&22	\\
CPD~-59~2635A	&O8~V		&4.96	&48.44	&0.26	&2100	&91	\\
CPD~-59~2635B	&O9.5~V		&4.68	&47.88	&0.15	&1660	&33	\\
CPD~-59~2636A	&O7~V		&5.14	&48.75	&0.4	&2320	&170	\\
CPD~-59~2636B	&O8~V		&4.96	&48.44	&0.26	&2100	&91	\\
CPD~-59~2636C	&O9~V		&4.77	&48.06	&0.18	&1820	&47	\\
CPD~-59~2641	&O5~V		&5.49	&49.22	&1.3	&2760	&780	\\
HDE~303311	&O5~V		&5.49	&49.22	&1.3	&2760	&780	\\
HDE~305536	&O9~V		&4.77	&48.06	&0.18	&1820	&47	\\
HD~93027	&O9.5~V		&4.68	&47.88	&0.15	&1660	&33	\\
HD~93028	&O9~V		&4.77	&48.06	&0.18	&1820	&47	\\
HD~93130	&O6~III(f)	&5.61	&49.32	&2.2	&2560	&1140	\\
HD~93146	&O6.5~V((f))	&5.23	&48.88	&0.6	&2460	&290	\\
HD~93160	&O6~III((f))	&5.61	&49.32	&2.2	&2560	&1140	\\
HD~93161Aa	&O8~V		&4.96	&48.44	&0.26	&2100	&91	\\
HD~93161Ab	&O9~V		&4.77	&48.06	&0.18	&1820	&47	\\
HD~93161B	&O6.5~V((f))	&5.23	&48.88	&0.6	&2460	&290	\\
HD~93206A	&O9.7~Ib:(n)	&5.57	&49.00	&1.9	&1510	&340	\\
HD~93206B	&O9~III		&5.21	&48.65	&0.5	&1820	&130	\\
HD~93222	&O7~III(f)	&5.48	&49.13	&1.5	&2320	&640	\\
HD~93343	&O7~V(n)	&5.14	&48.75	&0.4	&2320	&170	\\
HD~93403A	&O5~III(f)var	&5.73	&49.48	&3.4	&2760	&2050	\\
HD~93403B	&O7~V		&5.14	&48.75	&0.4	&2320	&170	\\
Tr16-2		&B1~V		&4.37	&47.28	&0.08	&1180	&8.8	\\
Tr16-3		&O8.5~V		&4.86	&48.27	&0.22	&1960	&67	\\
Tr16-4		&B1~V		&4.37	&47.28	&0.08	&1180	&8.8	\\
Tr16-5		&B1~V		&4.37	&47.28	&0.08	&1180	&8.8	\\
Tr16-8		&B0.5~V		&4.47	&47.50	&0.10	&1460	&17	\\
Tr16-9		&O9.5~V		&4.68	&47.88	&0.15	&1660	&33	\\
Tr16-10		&B0~V		&4.57	&47.70	&0.12	&1510	&22	\\
Tr16-11		&B1.5~V		&4.28	&47.05	&0.06	&960	&4.4	\\
Tr16-12		&B1~V		&4.37	&47.28	&0.08	&1180	&8.8	\\
Tr16-13		&B1~V		&4.37	&47.28	&0.08	&1180	&8.8	\\
Tr16-14		&B0.5~V		&4.47	&47.50	&0.10	&1460	&17	\\
Tr16-16		&B1~V		&4.37	&47.28	&0.08	&1180	&8.8	\\
Tr16-17		&B1~V		&4.37	&47.28	&0.08	&1180	&8.8	\\
Tr16-19		&O9.5~V		&4.68	&47.88	&0.15	&1660	&33	\\
Tr16-20		&B1~V		&4.37	&47.28	&0.08	&1180	&8.8	\\
Tr16-21		&O8~V		&4.96	&48.44	&0.26	&2100	&91	\\
Tr16-22		&O8.5~V		&4.86	&48.27	&0.22	&1960	&67	\\
Tr16-23		&O7~V		&5.14	&48.75	&0.4	&2320	&170	\\
Tr16-24		&B2~V		&4.19	&46.80	&0.05	&750	&2.2	\\
Tr16-27		&B1~V~?		&4.37	&47.28	&0.08	&1180	&8.8	\\
Tr16-28		&B2~V		&4.19	&46.80	&0.05	&750	&2.2	\\
Tr16-29		&B2~V		&4.19	&46.80	&0.05	&750	&2.2	\\
Tr16-31		&B0.5~V		&4.47	&47.50	&0.10	&1460	&17	\\
Tr16-37		&B1~V~?		&4.37	&47.28	&0.08	&1180	&8.8	\\
Tr16-74		&B1~V		&4.37	&47.28	&0.08	&1180	&8.8	\\
Tr16-76		&B2~V		&4.19	&46.80	&0.05	&750	&2.2	\\
Tr16-94		&B1.5~V		&4.28	&47.05	&0.06	&960	&4.4	\\
Tr16-115	&O8.5~V		&4.86	&48.27	&0.22	&1960	&67	\\
Tr16-122	&B1.5~V		&4.28	&47.05	&0.06	&960	&4.4	\\
Tr16-124	&B1~V		&4.37	&47.28	&0.08	&1180	&8.8	\\
Tr16-126	&O9~V		&4.77	&48.06	&0.18	&1820	&47	\\
Tr16-127	&O9~V		&4.77	&48.06	&0.18	&1820	&47	\\
Tr16-245	&B0~V		&4.57	&47.70	&0.12	&1510	&22	\\
%
\hline
\end{tabular}

Note --- $\eta$ Car's hypothetical value for $Q_H$ on the main
sequence is taken by adopting the same $Q_H$/L$_{Bol}$ ratio as for
HD~93129Aa.  References. --- Nelan et al.\ (2004); Massey \& Johnson
(1993); Walborn (1973, 1995; 2002, 2005); Penny et al.\ (1993);
Repolust et al.\ (2004); Naz\'{e} et al.\ (2005; for HD93161); Smith
et al.\ (1996); Crowther et al.\ (1995, 2002; although $Q_H$ and
$\dot{M}$ values for the WNL stars are from Crowther 2005, private
comm., modified to include clumping in the wind).

\end{minipage}
\end{table*}

\section{STELLAR CONTENT}

Tables 2--6 compile the stellar content powering the Carina Nebula.
References for the stellar content of each cluster are noted in each
table.  Several reviews exist in the literature (especially Walborn
1995, 2005; Feinstein 1995); these were invaluable in compiling the
tables.  Every effort has been made to provide the most recent
information with regard to binarity/multiplicity through spectral
types or through high spatial resolution techniques, especially for
Tr14 and 16.  This, of course, does not preclude the possibility that
additional companions may be identified, and their influence should be
considered by readers in that case.

Following Walborn (1995) and Tapia et al.\ (1988), it is assumed that
all these clusters share a common distance to the Carina Nebula, taken
here to be 2.3$\pm$0.1 kpc (Smith 2002$a$; Walborn 1995; Allen \&
Hillier 1993).  The distance of 2.3 kpc determined for $\eta$ Carinae
itself is very reliable, having been derived from proper motions and
Doppler velocities of the expanding Homunculus (Smith 2002$a$; Allen
\& Hillier 1993). Furthermore, $\eta$ Car is known to be at the same
distance as the Keyhole nebula, since its reflected spectrum is seen
across the Keyhole (Walborn \& Liller 1977; Lopez \& Meaburn 1986).
Differences in photometric distance moduli given in the literature are
probably due to anomalous or varying reddening laws, patchy
extinction, small age differences, or other factors.  For example,
photometric distances as high as 3.7 kpc have been given for Tr15
(Walborn 1973), requiring that Tr15 be several hundred pc {\it behind}
the rest of the Carina nebula.  This seems implausible, although not
impossible, for several reasons, including the fact that Tr15 is seen
along the same line of sight as dense gas and dust structures, some of
which are dust pillars that point toward Tr16.  Yet, stars in Tr15
have an average $E(B-V)$=0.5 (Feinstein et al.\ 1980) -- roughly the
same as Tr14 and 16 (Walborn 1995; Feinstein et al.\ 1973) -- which
could not be the case if Tr15 were far behind the rest of the nebula
and obscured.  Similarly, relatively large distances to Tr14 (Vazquez
et al.\ 1996) are implausible.  Even though Tr14 must be somewhat more
distant than Tr16 (perhaps as much as 20 pc) because silhouette
objects projected in front of it point toward Tr16 (Smith et al.\
2003$a$), it cannot be several hundred parsecs farther away.  The
outermost filaments of the Carina region span a distance of $\sim$120
pc (Smith et al.\ 2000), yet Tr14 and 16 are projected only about 5 pc
from each other on the sky, and even the Keyhole Nebula near Tr16 shows
signs of ionization from the northwest (i.e. from Tr14).

Additional clusters that are sometimes associated with Carina, such as
NGC 3324 and NGC 3293, are ignored in the present analysis.  They do
not appear to lie within the nebular structures that define the
outermost boundaries of the Carina Nebula, and the stars of NGC~3324
are associated with their own separate circular H~{\sc ii} region.

\subsection{Trumpler 16}

Energy input in Carina is dominated by the massive cluster Tr 16, of
which the infamous massive star $\eta$ Car is the most luminous
member.  This is clear even before a detailed look at its stellar
content, since optical images of Carina show numerous sculpted dust
pillars and other elongated structures located across several degrees
of the sky, and nearly all of them point back toward $\eta$ Car and
the center of Tr16 (Smith et al.\ 2000, 2003$a$, 2004, 2005; Walborn
2002).

As noted by Walborn (1995) and others, Cr 228 and Cr 232 are probably
part of Tr16, where their apparent separation on the sky is an
artifact caused by dust lanes on the near side of the nebula.  Thus,
Table 2 includes Cr 228 and 232 as members of a larger Tr16 cluster.
Table 2 also includes the O5 V star HDE 303311, which is part of
Cr232, as a member of Tr16, even though its membership is somewhat
uncertain (Walborn 1995).  In this and other cases, whether this star
is considered part of Tr16 or Tr14 has no net effect on the global
energetics of the region, but it may inflate the relative importance
of Tr16.

Also included within Tr16 in Table 2 are the three WNL stars in the
region.  HD~93162 is almost certainly a member of Tr16.  The two other
WNL stars, HD~92740 and HD~93131, are quite a bit farther away from
$\eta$ Car on the sky.  Walborn (1995) notes that they are both
roughly equidistant ($\sim$20 pc) from $\eta$ Car, and that they could
easily have traveled this far in 2 Myr with a kick of $\sim$10 km
s$^{-1}$.  Previously, on the main-sequence, they presumably had O2
spectral types like HD~93129A, and somewhat different $Q_H$ values.
In order to account for this stellar evolution, the cumulative
``main-sequence'' values in Table 7 (see \S 4) were calculated using
$Q_H$ values lowered by 0.1 dex for each WNL star.  This is roughly
correct if they were $\sim$0.1 dex less luminous on the main-sequence
(Maeder \& Meynet 2000), with the same $Q_H$/L$_{Bol}$ ratio as
HD93129A.

The age of Tr16 is probably about 2-3 Myr.  This comes from the fact
that it contains H-rich WNL stars but no He-rich WR stars, and that
$\eta$ Car has a likely ZAMS mass $>$100 M$_{\odot}$ and has already
evolved off the main-sequence --- as well as the fact that there is no
clear evidence for a supernova having exploded yet.

When considering the cumulative effect of the massive stars in Tr16,
we must bear in mind the peculiar present state of its most luminous
member, $\eta$ Car.  As we observe it today, $\eta$ Car is a pitiful
source of UV photons for the larger region.  Whatever ionizing
radiation is able to escape its very dense optically-thick stellar
wind\footnote{The mass-loss rate for $\eta$~Car listed in Table 2 is
from Hillier et al.\ (2001), and includes the effects of clumping in
the wind.} (Hillier et al.\ 2001; Smith et al.\ 2003$b$) is eventually
absorbed by dust in the surrounding Homunculus nebula and converted
into infrared radiation (Smith et al.\ 2003$c$; Cox et al.\ 1995).
However, more than 10$^4$ years ago, before it entered its
presently-observed luminous blue variable (LBV) phase and was on the
main sequence --- or even more than 160 years ago before the
Homunculus existed --- $\eta$ Car was a {\it much} stronger source of
UV radiation.  Indeed, it probably contributed $\sim$20\% of the total
ionizing photons for the entire region (see below).  By the same
token, its main-sequence stellar wind was weaker than its
presently-observed LBV wind.  Its spectral type prior to the LBV phase
is of course unknown, but given its extreme luminosity, it is
reasonable to assume that it was an O2 supergiant like HD~93129A while
on the main sequence.  From stellar evolution tracks for very massive
stars (e.g., Young 2005; Maeder \& Meynet 2000), it is likely that
$\eta$ Car was $\sim$0.1 dex less luminous at that time.  The value
for $Q_H$ at that time is then calculated assuming this lower
L$_{Bol}$, and assuming the same ratio of $Q_H$/L$_{Bol}$ as for
HD93129A.

It is also likely that $\eta$~Car has a close companion star in a 5.5
yr eccentric orbit (Damineli et al.\ 2000; Corcoran 2005; Duncan et
al.\ 1995).  While its UV radiation does not escape the Homunculus at
the present time, it did contribute to the ionization of the larger
region throughout the 2-3 Myr lifetime of Tr16.  The nature of $\eta$
Car's companion star is highly uncertain and a matter of current
debate (Pittard \& Corcoran 2002; Ishibashi et al.\ 1999; Davidson
1999; Duncan et al.\ 1995); a tentative spectral type of O5 V is
adopted in Table 2.  The bolometric luminosity of $\eta$ Car (the
primary star) in Table 2 has the contribution of its companion
subtracted.

%
%
\begin{table*}\begin{minipage}{4.75in}
\caption{Stellar Content of Trumpler 14}
\scriptsize
\begin{tabular}{@{}llccccc}\hline\hline
 Name &Sp. &log L &log $Q_H$ &$\dot{M}$ &V$_{\infty}$ &$L_{SW}$ \\
 \ &Type &(L$_{\odot}$) &(s$^{-1}$) &(10$^{-6}$ M$_{\odot}$ yr$^{-1}$) 
         &(km s$^{-1}$) &(L$_{\odot}$) \\ \hline
HD~93129Aa	&O2~If*		&6.17	&49.96	&8.32	&3200	&6780	\\
HD~93129Ab	&O3.5~V		&5.76	&49.54	&2.5	&3080	&1870	\\
HD~93129B	&O3.5~V((f+))	&5.76	&49.54	&2.5	&3080	&1870	\\
HD~93128	&O3.5~V((f+))	&5.76	&49.54	&2.5	&3080	&1870	\\
CPD-58~2611	&O6~V((f))	&5.32	&48.99	&0.8	&2560	&410	\\
CPD-58~2620	&O6.5~V((f))	&5.23	&48.88	&0.6	&2460	&290	\\
Tr14-3		&B0.5~IV-V	&4.50	&47.70	&0.15	&1460	&26	\\
Tr14-4		&B0~V		&4.57	&47.70	&0.12	&1510	&22	\\
Tr14-5		&O9~V		&4.77	&48.06	&0.18	&1820	&47	\\
Tr14-6		&B1~V		&4.37	&47.28	&0.08	&1180	&8.8	\\
Tr14-9		&O8~V		&4.96	&48.44	&0.26	&2100	&91	\\
Tr14-18		&B0~V		&4.57	&47.70	&0.12	&1510	&22	\\
Tr14-21		&O9~V		&4.77	&48.06	&0.18	&1820	&47	\\
Tr14-27		&O9~V		&4.77	&48.06	&0.18	&1820	&47	\\
Tr14-30		&B0~III-IV	&4.90	&48.10	&0.25	&1510	&45	\\
%
\hline 
\end{tabular}

References. --- Nelan et al.\ (2004); Massey \& Johnson (1993);
Walborn (1973, 1995; 2005); Penny et al.\ (1993); Repolust et al.\
(2004).

\end{minipage}
\end{table*}

\subsection{Trumpler 14}

%
%
\begin{table*}\begin{minipage}{4.75in}
\caption{Stellar Content of Trumpler 15}
\scriptsize
\begin{tabular}{@{}llccccc}\hline\hline
 Name &Sp. &log L &log $Q_H$ &$\dot{M}$ &V$_{\infty}$ &$L_{SW}$ \\
 \ &Type &(L$_{\odot}$) &(s$^{-1}$) &(10$^{-6}$ M$_{\odot}$ yr$^{-1}$) 
         &(km s$^{-1}$) &(L$_{\odot}$) \\ \hline
HDE~303304	&O8~V		&4.96	&48.44	&0.26	&2100	&91	\\
HD~93249	&O8~II/O9~III	&5.28	&48.75	&0.7	&1960	&210	\\
HD~93342	&O9~III		&5.21	&48.65	&0.5	&1820	&130	\\
Tr15-18		&O9~I-II	&5.61	&49.11	&2.5	&1660	&540	\\
Tr15-2		&O9~III		&5.21	&48.65	&0.5	&1820	&130	\\
Tr15-19		&O9~V		&4.77	&48.06	&0.18	&1820	&47	\\
HD~93190	&B0~IV		&4.83	&48.00	&0.21	&1510	&38	\\
Tr15-3		&B2~V		&4.19	&46.80	&0.05	&750	&2.2	\\
Tr15-4		&B1~V		&4.37	&47.28	&0.08	&1180	&8.8	\\
Tr15-7		&B2~V		&4.19	&46.80	&0.05	&750	&2.2	\\
Tr15-9		&B1~V		&4.37	&47.28	&0.08	&1180	&8.8	\\
Tr15-10		&B2~V		&4.19	&46.80	&0.05	&750	&2.2	\\
Tr15-13		&B1~V		&4.37	&47.28	&0.08	&1180	&8.8	\\
Tr15-14		&B2~V		&4.19	&46.80	&0.05	&750	&2.2	\\
Tr15-15		&B0.5~IV-V	&4.47	&47.50	&0.10	&1460	&17	\\
Tr15-20		&B1~V ?		&4.37	&47.28	&0.08	&1180	&8.8	\\
Tr15-21		&B0~III		&5.08	&48.28	&0.3	&1510	&54	\\
Tr15-23		&B0~V		&4.57	&47.70	&0.12	&1510	&22	\\
Tr15-26		&B0.5~V		&4.47	&47.50	&0.10	&1460	&17	\\
%
\hline
\end{tabular}

References. --- Feinstein et al.\ (1980); Morrell et al.\ (1988).

\end{minipage}
\end{table*}


Trumpler 14 is a smaller, much more compact, and perhaps somewhat
younger cluster than Tr16.  Some members of Tr16, like HD~93161,
93160, and 93250, are projected nearby on the sky (Walborn 1973) and
are often mistaken as members of Tr14.  As noted above, it is likely
that Tr14 is perhaps 10--20 pc more distant than Tr16, since
silhouette objects seen in front of it point toward Tr16 (Smith et
al.\ 2003), but it is still part of the central engine that powers the
Carina Nebula.  Its most conspicuous member is the prototypical O2~If*
supergiant HD~93129A.

Because of its compact size, its relatively faint upper main sequence,
and certain spectral diagnostics, Tr14 may be 1-2 Myr younger than
Tr16 (Walborn 1995).  Penny et al.\ (1993) estimate ages younger than
1 Myr for HD~93129A and HD~93128, while Vazquez et al.\ (1996) find a
cluster age of $\sim$1.5 Myr.  In addition, Tr14 remains closer to its
natal molecular cloud than Tr16 (e.g., Brooks et al.\ 2003).

The stellar content summarized in Table 3 is compiled from several
references listed at the end of the table, with particular attention
to recent results on the binarity and multiplicity of stars in the
cluster (see Nelan et al.\ 2004; Walborn 2005), including a new
companion to HD~93129A (Nelan et al.\ 2004).  The greatest source of
uncertainty in the net output of this cluster is in the parameters for
HD93129A itself.  The value of log L$_{Bol}$=6.17 is adopted from
Repolust et al.\ (2004), and the value for $Q_H$ was calculated from
the luminosity using the revised $T_{eff}$ calibration scale of
Martins et al.\ (2005).

%
%
\begin{table*}\begin{minipage}{4.5in}
\caption{Stellar Content of Bochum 10}
\scriptsize
\begin{tabular}{@{}llccccc}\hline\hline
 Name &Sp. &log L &log $Q_H$ &$\dot{M}$ &V$_{\infty}$ &$L_{SW}$ \\
 \ &Type &(L$_{\odot}$) &(s$^{-1}$) &(10$^{-6}$ M$_{\odot}$ yr$^{-1}$) 
         &(km s$^{-1}$) &(L$_{\odot}$) \\ \hline
HD~92809	&WC6		&5.32	&49.20	&16	&2280	&6620	\\
HD~92725	&O9.5~III	&5.15	&48.42	&0.4	&1660	&87	\\
HD~92964	&B2~Ia		&5.30	&48.35	&0.5	&750	&22	\\
HD~92759	&B0~III		&5.08	&48.28	&0.3	&1510	&54	\\
HD~92894	&B0~IV		&4.83	&48.00	&0.21	&1510	&38	\\
HD~93002	&B0~IV		&4.83	&48.00	&0.21	&1510	&38	\\
HDE~302989	&B2~V		&4.19	&46.80	&0.05	&750	&2.2	\\
HDE~303190	&B1~V		&4.37	&47.28	&0.08	&1180	&8.8	\\
HDE~303296	&B1~V		&4.37	&47.28	&0.08	&1180	&8.8	\\
HDE~303297	&B1~V		&4.37	&47.28	&0.08	&1180	&8.8	\\
Bo10-2		&B0.5~IV	&4.74	&47.80	&0.18	&1460	&30	\\
Bo10-7		&B2~IV		&4.49	&47.12	&0.08	&750	&3.5	\\
Bo10-15		&B1~V		&4.37	&47.28	&0.08	&1180	&8.8	\\
\hline
\end{tabular}

References. --- Feinstein (1981); Fitzgerald et al.\ (1987); Prinja et
al.\ (1990); Smartt et al.\ (2001).

\end{minipage}
\end{table*}

%
%
\begin{table*}\begin{minipage}{4.5in}
\caption{Stellar Content of Bochum 11}
\scriptsize
\begin{tabular}{@{}llccccc}\hline\hline
 Name &Sp. &log L &log $Q_H$ &$\dot{M}$ &V$_{\infty}$ &$L_{SW}$ \\
 \ &Type &(L$_{\odot}$) &(s$^{-1}$) &(10$^{-6}$ M$_{\odot}$ yr$^{-1}$) 
         &(km s$^{-1}$) &(L$_{\odot}$) \\ \hline
HD~93632	&O4-5~III(f)	&5.79	&49.56	&4.0	&2870	&2600	\\
Bo11-2		&O9~V		&4.77	&48.06	&0.18	&1820	&47	\\
Bo11-3		&B2~V		&4.19	&46.80	&0.05	&750	&2.2	\\
Bo11-4		&B0.5~V		&4.47	&47.50	&0.10	&1460	&17	\\
Bo11-5		&O9~V		&4.77	&48.06	&0.18	&1820	&47	\\
Bo11-9		&O9~IV		&4.99	&48.36	&0.34	&1820	&89	\\
Bo11-10		&O9~IV		&4.99	&48.36	&0.34	&1820	&89	\\
%
\hline
\end{tabular}

References. --- Fitzgerald et al.\ (1987); Walborn (1973).

\end{minipage}
\end{table*}


\subsection{Trumpler 15}

Tr15 is an open cluster located about 20\arcmin\ north of Tr16 on the
sky.  It is thought to be at the same distance and has roughly the
same reddening as Tr16, but it is somewhat older, with a likely age of
6$\pm$3 Myr (Feinstein et al.\ 1980; Morrell et al.\ 1988).  It is
seen near a very bright red star on the sky -- the M2 Ia supergiant RT
Car, which is probably not a member of the cluster.  Table 4 adopts
the cluster members listed by Feinstein et al.\ (1980) and Morrell et
al.\ (1988), including the O8 V star HD~303304 which was listed as a
possible nonmember by Feinstein et al.\ (1980), as well the O9 giant
HD~93342 and the highly-reddened O9 I-II star Tr15-18, which were
listed as possible members.  Table 4 does not include the non-member
O5 III star HD~93403, which is taken to be a member of Tr16.

\subsection{Bochum 10}

Bo10 is a relatively meager cluster located roughly 40\arcmin\
northwest of Tr16, with a handful of OB stars and a probable age of
$\sim$7 Myr (Feinstein 1981; Fitzgerald et al.\ 1987).  Thus, it may
be significantly older than Tr 14 and 16, but perhaps not much older
than Tr15.  Feinstein found a large distance of 3.6 pc, but a later
analysis by Fitzgerald et al.\ (1987) favored a closer distance of 2.5
kpc, placing it inside the Carina complex associated with some nearby
nebulosity.  Because of its larger age of $\sim$7 Myr, it is likely
that the nearby WR star HD~92809 may be associated with Bo10, or at
least with the larger Carina complex.  Therefore, HD~92809 is included
in Table 5 along with the other members of Bo10.  For the total FUV
luminosity of HD92809 in Table 7 (see \S 4), we adopt
L$_{FUV}$/L$_{Bol}$=0.39 (for 912--3650 \AA) following the study of
Smartt et al.\ (2001).

\subsection{Bochum 11}

Bo11 is a loose open cluster in the southeastern Carina Nebula.
Fitzgerald et al.\ (1987) find an age for Bo11 of less than 3 Myr.
They also find that the cluster includes several pre-main-sequence
stars, consistent with an age as young as 0.3 Myr.  This very young
age is interesting in light of the fact that Bo11 is located amid what
is currently the most active region of star formation in Carina -- the
so-called South Pillars (Smith et al.\ 2000; Rathborne et al.\ 2004).
The age and stellar content is similar to the nearby embedded cluster
known as the Treasure Chest (Smith et al.\ 2005; H\"{a}gele et al.\
2004).\footnote{Although it is not discussed in detail here, the O9.5
V star CPD-59$\arcdeg$2661 in the Treasure Chest cluster is included
in Table 7.}  There is considerable discrepancy as to the spectral
type of the most luminous star in Bo 11, which is HD 93632 or Bo11-1.
Fitzgerald et al.\ (1987) list this star as O8 Iap, whereas Walborn
(1973) classifies it as O4-5 III(f).  Walborn (1973) noted that the
spectral type was somewhat variable, while Fitzgerald et al.\ (1987)
noted a few problems with their photometric classification of stars in
Bo11.  A spectral type of O4-5 III(f) is adopted in Table~6.

\section{CUMULATIVE EFFECTS}

%
\begin{table*}\begin{minipage}{4.75in}
\caption{Total Stellar Energy Input}
\begin{tabular}{@{}lcccccc}\hline\hline
Cluster &Number of &log L &log $Q_H$ &log L(FUV) &$\dot{M}$ &$L_{SW}$ \\
  &O stars &(L$_{\odot}$) &(s$^{-1}$) &(L$_{\odot}$) 
  &(10$^{-6}$ M$_{\odot}$ yr$^{-1}$)  &(L$_{\odot}$) \\ \hline
Tr16 (MS)	&47	&7.215	&50.91	&6.91	&91	&45400	\\
Tr16 (LBV)	&43	&7.240	&50.78	&7.05	&1083	&67000	\\
Tr16 (now)	&42	&7.240	&50.77	&6.79	&1083	&67000	\\
Tr14		&10	&6.61	&50.34	&6.31	&18.7	&13500	\\
Tr15		&6	&6.18	&49.56	&5.88	&5.9	&1300	\\
Bo10		&1	&6.00	&49.42	&5.69	&18.3	&7120	\\
Bo11		&5	&6.00	&49.64	&5.70	&5.2	&2900	\\
CPD-59$^{\circ}$2661 &1	&4.68	&47.88	&4.38	&0.15	&33	\\ \hline
TOTAL (MS)	&70	&7.38	&51.06	&7.08	&139	&70200	\\
TOTAL (LBV)	&66	&7.40	&50.97	&7.18	&1131	&91800	\\
TOTAL (now)	&65	&7.40	&50.96	&7.00	&1131	&91800	\\
%
\hline
\end{tabular}\end{minipage}
\end{table*}

Table 7 lists the total luminosity, ionizing flux, total FUV (Balmer
continuum) luminosity, mass loss, and mechanical luminosity for each
cluster, as well as the cumulative total of all clusters for each
parameter.  For the FUV luminosity (in all cases except where noted
above), L$_{\rm FUV} \simeq$0.5L$_{\rm Bol}$ is assumed, which is
adequate considering the inherent uncertainty in the bolometric
luminosity as a function of spectral type.  Brooks et al.\ (2003)
found a similar result for the FUV luminosity of Tr14 alone.  More
detailed stellar stmosphere models typically have L$_{FUV}$/L$_{Bol}$
values of 0.39--0.85 for 912--3650 \AA\ (e.g., Smartt et al.\ 2001;
Crowther et al.\ 2002; Crowther 2005, private comm.).  Three cases are
given for Tr16 and for the total of all clusters:

The first case correponds to the history of Carina up until recent
times, when $\eta$~Car and its companion were not surrounded by a dust
shell and did not have a dense LBV wind choking off the Lyman
continuum flux, and when the WNL stars in Tr16 were O2 stars.  For
this first case, there was a total of 70 O-type stars in Carina,
producing $Q_H$=1.15$\times$10$^{51}$ s$^{-1}$.  This would be the
appropriate number to adopt when considering the history and formation
of the nebula, the lifetimes of evaporating proplyds, globules, and
dust pillars, triggered star formation by radiative driven implosion,
and the growth of the cavity that will blow out of the Galactic plane
as a bipolar superbubble.

The second case is the LBV phase of $\eta$ Car during the last
$\sim$10$^4$ yr, when $\eta$ Car had a dense wind that reprocessed
most of the Lyman continuum radiation into the Balmer continuum (FUV).
During this phase, $\eta$~Car was not surrounded by a dense dust
shell, allowing its FUV luminosity to escape.  Since the UV radiation
of its companion could also escape, and since the WNL stars were no
longer on the main sequence, the total number of O stars is listed as
66 in Table 7.  This brief phase may be important in understanding the
recent photoevaporation of neutral globules and photodissociation
regions throughout the nebula, since $\eta$ Car was such an incredibly
luminous FUV source.

The third case represents the currently-observed state of the Carina
Nebula, when $\eta$~Car and its companion are surrounded by an
obscuring dust shell, blocking all their contribution to the total
ionizing flux and FUV luminosity.  With $\eta$~Car's companion blocked
by the Homunculus, the effective number of O stars is reduced to 65.
For this third case, the cumulative ionization source is
$Q_H$=9$\times$10$^{50}$ s$^{-1}$.  This would be the appropriate
number to adopt when considering the current UV flux incident upon
evaporating proplyds, globules, and irradiated jets over most of the
nebula.

For most of its lifetime, gas and dust in the Carina Nebula has been
exposed to an ionizing luminosity about 150 times stronger than that
of the Orion nebula.  This changed 160 years ago when $\eta$~Car
ejected a thick dust shell that cut off essentially all of its UV
output, and the total Q$_H$ of the Carina Nebula dropped by about 20\%
due to the loss of ionization from the region's most luminous
member.\footnote{Actually, the change might not have been so sudden,
since much of the Lyman continuum flux from $\eta$~Car may have
already dropped when it developed a thick stellar wind upon entering
the LBV evolutionary phase in the last 10$^4$ yr.  During this time,
the majority of $\eta$ Car's radiation escaped as FUV photons.}

This change illustrates the dominant role of the most luminous star in
any region, and highlights a truly unique property of the Carina
Nebula --- it offers a laboratory where we can address variability in
Q$_H$ over short timescales.  This behavior has not been witnessed
anywhere else in our Galaxy to a similar degree.  We can observe some
of the effects of this variable UV output --- for example, neutral
cometary globules and dust pillars that point at $\eta$~Car are seen
only in silhouette today, but their shape suggests that they were
formed by a strong UV flux from $\eta$ Car in the past (Smith et al.\
2003$a$).  It is interesting to note that the outer edges of the
Carina Nebula, beyond a radius of 160 light yr or 50 pc
(1.5$\times$10$^{20}$ cm) from $\eta$~Car, have not yet seen the Great
Eruption of $\eta$~Car, and still see its pre-eruption UV output.
Also, if the complex nested circumstellar ejecta around $\eta$~Car are
a fair indication, this type of UV cutoff has probably happened
multiple times in the past due to previous outbursts, and may yet
happen again.

The main uncertainty in deriving the pre-eruption UV output of Carina
is in the spectral type and ionizing luminosity of $\eta$~Car itself,
which has a likely uncertainty of perhaps $\pm$15\% (roughly 3\% of
the total for the region) if the models at lower luminosity and later
spectral types are correct.  In terms of the cumulative effect of the
entire stellar population, then, dwarf stars later than O9 can be
safely ignored, since their collective contribution is less than the
uncertainty in the value for Q$_H$ from the most luminous members.
For the other luminous members, in Case 1 (when $\eta$ Car was on the
main sequence), about 8\% of the total Q$_H$ came from the O2
supergiant HD~93129Aa alone, 15\% came from the 3 WNL stars, and 15\%
from the remaining five O3.5~V stars.  For case 3, corresponding to
the present state of affairs, these contribution have changed to about
10\% of the total from HD~93129Aa alone, 23\% from the 3 WNL stars,
and 19\% from the remaining five O3.5V stars.

\begin{figure}\begin{center}
\epsfig{file=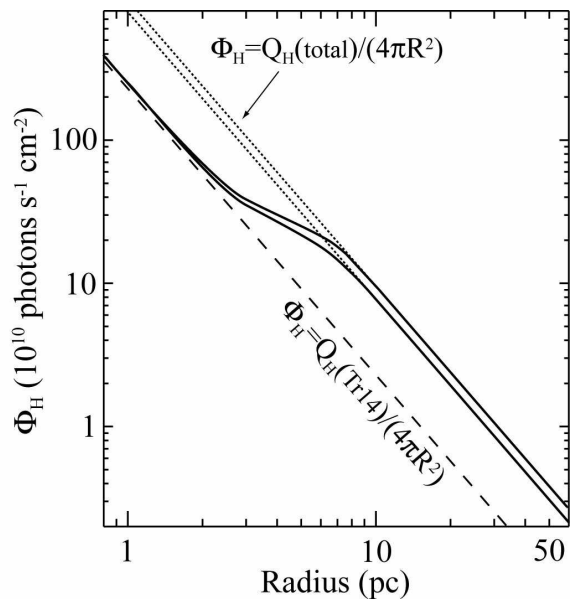,width=3.0in}
\end{center}
\caption{Ionizing photon flux $\Phi$ as a function of radius.  The two
solid curves are for Case 1 and 3 (i.e. $\eta$ Car on the main
sequence and at the present time).}
\end{figure}

Figure 1 shows the ionizing photon flux $\Phi_H$ as a function of
radius.  The two solid curves are for Case 1 and 3 (i.e. $\eta$ Car on
the main sequence and at the present time).  Beyond a radius of
roughly 10 pc from the center of the nebula, the ionizing source can
be treated roughly as a point source, so that the photon flux drops as
r$^{-2}$.  Within a few parsecs of any cluster, however, the local
radiation from that cluster becomes more important than the diminished
collective effects from all clusters.  This is due to the fact that
the stellar population in Carina is a loose aggregate, rather than a
single dense cluster.  The specific case of Tr14 is shown in Figure 1,
but within a few parsecs of any individual cluster, its own ionization
will dominate, while the influence of more distant clusters
diminishes.

\begin{figure}\begin{center}
\epsfig{file=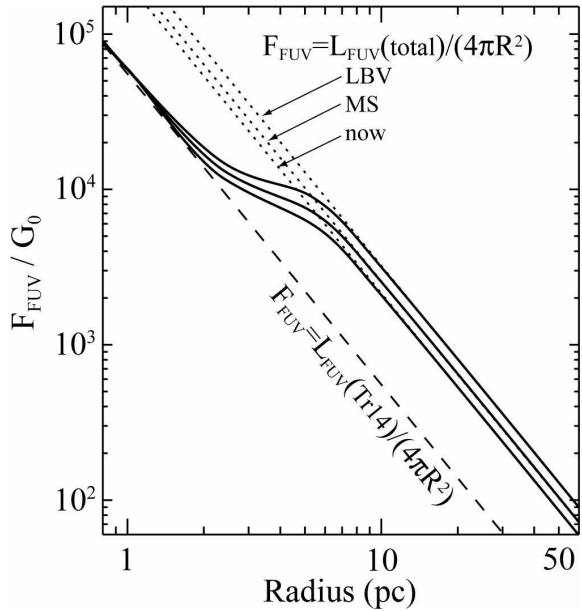,width=3.0in}
\end{center}
\caption{Same as figure 1, but for FUV radiation, plotted in units of
the Habing flux, $G_0$=1.6$\times$10$^{-3}$ ergs s$^{-1}$ cm$^{-2}$.
The three cases from Table 7 are shown.}
\end{figure}

Figure 2 shows the radial dependence of the FUV radiation field in the
Carina Nebula, compared to the Habing flux, $G_0$=1.6$\times$10$^{-3}$
ergs s$^{-1}$ cm$^{-2}$ (Habing 1968).  The three curves correspond to
$\eta$ Car on the main sequence, in the LBV phase, and in its present
state obscured by its own dust shell.  The strongest FUV fluxes in the
region arise when the LBV phase of $\eta$ Car is considered, when most
of its bolometric luminosity escaped in the Balmer continuum.
Throughout the Carina Nebula, this resulted in a local FUV flux that
was roughly 75\% higher than the present state when $\eta$ Car is
surrounded by dust.  Thus, for extremely massive stars, their changing
UV output in various evolutionary states can profoundly effect the
radiation field of their environments, and consequently, the
evaporation of protoplanetary disks and the erosion of dust pillars.
Again, however, individual clusters dominate over the collective
radiation field within a few parsecs of any cluster core.

The present total mechanical luminosity from stellar winds is just
under 10$^5$ $L_{\odot}$, or 7$\times$10$^4$ $L_{\odot}$ during the
main-sequence phase, having contributed a total energy of about
2.6$\times$10$^{52}$ ergs so far during the 3 Myr lifetime of the
region.  This is higher than a previous estimate of several
$\times$10$^{51}$ ergs (Walborn 1982) because it includes the
influence of many more stars.  The estimated kinetic energy of the
expanding proto-superbubble (Smith et al.\ 2000; Walborn 1982) is only
5--30\% of this mechanical energy from stellar winds, suggesting that
most of the energy injected by stellar winds is radiated away, even in
the early phases of a bubble's growth.  The total energy radiated by
early-type stars during the same time was over 350 times more than the
mechanical energy.  This situation will change dramatically when the
most massive stars begin to explode as supernovae in the next few
10$^5$--10$^6$ yr, each injecting an additional 10$^{51}$--10$^{52}$
ergs in a very short time.

With a total mass-loss rate of $\dot{M} \approx 1.4\times 10^{-4} \
M_{\odot}$ yr$^{-1}$, the mass ejected by stellar winds over the past
3 Myr is about 420 $M_{\odot}$.  This is only about 10\% of the mass
of ionized gas estimated to be filling the interior of the H~{\sc ii}
region (e.g., Walborn 2005).  Thus, the majority of ionized gas inside
the H~{\sc ii} region cavity probably comes from photoevaporative
flows (e.g., Bertoldi 1989) off the surfaces of molecular clouds and
evaporating globules embedded within the region.  Note that the
present value of $\dot{M}$ for $\eta$ Car's brief LBV wind phase is
much greater than the cumulative mass loss of rest of the stars in
Carina combined.

\section{COMPARISON WITH OTHER MASSIVE CLUSTERS}

Finally, it is interesting to see how the stellar energy input in
Carina stacks up to other young massive star clusters.  As noted
earlier, in general, Carina's overall stellar content is similar to
the most well-studied massive clusters like NGC3603, Arches, and 30
Dor, except that these clusters are more compact than Carina.  For
example, Crowther \& Dessart (1998) list log $Q_H$=51.19 and
L$_{SW}$=1.5$\times$10$^5$ L$_{\odot}$ for NGC3603.  The total energy
input in Carina is essentially the same as these values, because
Crowther \& Dessart used an older T$_{eff}$ scale, leading to higher
$Q_H$ values for O stars, and did not account for wind clumping in
their analysis; if those same assumption were made here for Carina,
the results would be identical to NGC3603 within the uncertainties.
Carina has about 25--30\% of the total ionizing flux of the more
massive Arches cluster in the Galactic center (Figer et al.\ 2002),
and roughly 25\% of both the total $Q_H$ and L$_{SW}$ of R136 in 30
Doradus given by Crowther \& Dessart (1998).  Given the similar
cumulative energy input and stellar content, it is interesting that
Carina's clusters are less tightly bound than the others, especially
if this gives rise to different effects on the nearby star-forming
environment.  A future paper will investigate how the cumulative
stellar energy input assessed directly from the observed massive stars
in Carina compares with more indirect tracers of the energy budget of
the larger Carina Nebula.

\smallskip\smallskip\smallskip\smallskip
\noindent {\bf ACKNOWLEDGMENTS}
\smallskip
\scriptsize

I thank Paul Crowther for a detailed referee report and several
helpful suggestions regarding model parameters.  I also thank Nolan
Walborn and Kate Brooks for useful discussions and helpful
comments and corrections on the manuscript.  Support was provided by
NASA through grant HF-01166.01A from the Space Telescope Science
Institute, which is operated by the Association of Universities for
Research in Astronomy, Inc., under NASA contract NAS~5-26555.


\end{document}